\documentclass[useAMS,usenatbib,usegraphicx]{mn2e}

\title[The Spin Period of BF CYG]{Discovery of the 1.80 hr Spin Period of the White Dwarf of the Symbiotic System BF CYG}
\author[Liliana Formiggini and Elia M. Leibowitz ]
{Liliana Formiggini$^{1}$\thanks{E-mail:
lili@wise.tau.ac.il}
and Elia M. Leibowitz$^{1}$\thanks{E-mail: elia@wise.tau.ac.il}\\
$^{1}$The Wise Observatory and the School of Physics and Astronomy, Raymond
and Beverly Sackler Faculty of Exact Sciences \\ Tel Aviv University, Tel
Aviv 69978, Israel\\}
\begin{document}
\pagerange{\pageref{firstpage}--\pageref{lastpage}} \pubyear{2008}
\maketitle
\label{firstpage}
\begin{abstract}
We report on the discovery of a coherent periodicity  in the B light curve of
the symbiotic star BF Cyg. The signal was detected in some sections of the
light curve of the star recorded in the year 2003 as double hump periodic
variations with an amplitude of ~ 7 mmag. In the year 2004 the signal was
also present in only a subsection of the light curve. In that year, the
system was about twice as bright  and the amplitude of the oscillations was
about half of what it was in 2003. In 2004 the cycle structure was of a
single hump, the phase of which coincided with the phase of one of the
humps in the 2003 cycle. No periodic signal was detected in a third, short
series of observations performed in the year 2007, when the star was three
magnitudes brighter than in 2003. We interpret the periodicity as the spin
period of the white dwarf component of this interacting binary system. We
suggest that the signal in 2003 originated in two hot spots on  or near the surface
of the white dwarf, most likely around the two antipodes of an oblique
dipole magnetic field of this star. Magnetic field lines funneled accreted
matter from the wind of the cool component to the pole areas, where the
falling material  created the hot spots. This
process is apparently intermittent in its nature. In 2004, the activity
near only one pole was enhanced enough to raise the signal above the
threshold of our detection ability.
\end{abstract}


\section{Introduction-Rapid oscillation of  symbiotic systems}

 A typical configuration of a symbiotic system (SS) comprises a red giant,
a hot compact component, which in many cases is probably a white dwarf (WD)
and a surrounding nebula. Quite a few modes of variations take place in SSs, 
mainly on long time scales (Kenyon, 1986). 
One mode is  of cyclic variations, with periods  of  hundreds or  thousands of days, 
indicating the binary nature of the system. The cool giant in the system, can also 
show intrinsic variability such as radial pulsation  (Mira-type). 
Another type of variability is of an explosive character, in the form
of a single outburst, as for symbiotic novae, or multiple events.
Recent investigations, based on data assembled from many archival 
sources and historical data from old plate collections around the world,
discovered a periodicity on time scales of tens of years for the activity
events of some SSs. This periodicity, detected in  BF Cyg, YY Her and Z And  
(Formiggini \& Leibowitz, 1994; Leibowitz \& Formiggini, 2006--hereafter paper I,
Formiggini \& Leibowitz, 2006; Leibowitz \& Formiggini, 2008)
can be attributed to sun-like variability of the giant component of the
system.
Variability with time scales of the order of minutes similar to the
flickering typical of cataclysmic variables (CV) has been searched
for in SSs.  
Although there are few objects that are known to show 
flickering with amplitude of tens to hundreds of mmag (e.g.T CrB, CH Cyg,
RS Oph, {\it o} Ceti, MWC 560, RT Cru, V407 Cyg), most SSs show 
no detectable rapid optical variability down to a limit of a few 
mmag (Dobrzycka, Kenyon \& Milone, 1996, Sokoloski, Bildsten \& Ho 2001, 
Sokoloski \& Kenyon 2003, Gromadzki et al. 2006).
Only for one system, Z And, a 28 minutes periodic variation has been 
detected and explained as rotation of a magnetic white dwarf which is 
the hot component of the system (Sokoloski \& Bildsten 1999). This 
coherent oscillation was again detected by Gromadzki et al. (2006), but it 
was absent during the 2000-2002 fluctuation of Z And (Sokoloski et al. 2006).

Although  its mechanism in SSs is not yet well understood, 
the short-time  variability is believed to be related to the hot
component of the symbiotic binary system, namely the white dwarf
and/or the accretion disk which in some cases is likely to surround it.
However, the failure to detect flickering on most SSs may  be due  either to  
the absence of accretion disks or to observational constraints. For instance,
the emission of the nebula and/or nuclear burning material of the WD may 
reduce the amplitude of the variation and lower its detection probability.

We have started at the Wise Observatory of the Tel Aviv University an 
observational campaign on the short term variability of SSs. 
The aim of this research  is to target a few systems and observe them 
intensively in order to detect possible coherent short term variability in 
their LCs. Since the amplitude of the expected variability is of the order 
of few mmag, it is mandatory to collect a consistent amount of data in 
order to obtain a good precision.
Furthermore, the data analysis technique is of primary importance when
searching for such small amplitude variability phenomenon. Actually, the  
search of a periodic signal over the total length of the observations 
hypothesizes a coherence that cannot be assumed a priori.
We have developed  a MATLAB script that allows us to analyze the 
structure of the short time scales variability of SSs. This  analysis 
and the long time series of our observations, have indeed given a new 
insight into the structure of the variability of SSs stars.
 
In this paper we present the results of the observational campaign on one 
of our targets, the BF Cyg system.

\section{The symbiotic system BF Cyg}
 
We remind here that BF Cyg is one of the prototype symbiotic stars that was  
well observed at practically all frequencies of the electromagnetic spectrum 
(Mikolajewska et al. 1989; Fernandez-Castro et al. 1990; Gonz\'{a}les-Riestra et al. 1990).

The system consists of an M giant star and a hot compact star, possibly a WD, with 
an orbital period of 757.3 d. The nature of the hot component of BF Cyg is not known,
but from the optical and IUE ({\it International Ultraviolet Explorer}) spectra a 
temperature of 60000 K is inferred, as for a hot  white dwarf.

The historical LC of BF Cyg has been analysed in details in paper I and we 
remind here some of the findings. Its long term light curve (LC) shows a general 
decline, following its major outburst in 1894, as for a symbiotic nova system. 
Superposed on this decline, there are  several  brightenings of the 
system by one or more magnitudes, of a typical duration of a few years.
A periodicity in the occurrence of the outburst events of 6376 d has been 
discovered and interpreted as a sign of a dynamo generated  magnetic 
activity occurring in the cool giant component, similar to the solar cycle.
It was pointed out in paper I, however, that the magnetic 
cycle is far from being strictly periodic, as in the case of the sun.
The 6376 d  period detected in paper I, allowed the prediction of the
occurrence of an outburst event with maximum height in mid-year 2007,
around JD 2454236. 
Indeed, on 2006 July 31 the system entered into a new active phase which
continued during the years 2007 and 2008 (Munari et al. 2006,
Skopal et al. 2007). Up to now, the maximum light was reached around JD 2454724. 
When the ongoing outburst is completed and the system returns to its quiescent 
state, a better estimation of the mean period of the magnetic cycle 
can be obtained.
Another periodicity of 798.8 d is explained as the rotation period of the giant star
in the system (see paper I).

BF Cyg is one of 35 objects observed  by Sokoloski, Bildsten \& Ho (2001). 
While for most of these targets no variability was detected up to a limit of 
a few mmag, BF Cyg was classified among the four new candidate flickers, 
in need of further observation.
The system was observed by Sokoloski et al. (2001) three times in the B filter, 
near phases .89, .47 and .44 of its photometric cycle, where zero is the phase of 
the minimum light according to the ephemeris Min= JD 2415065 +757.3 $\times$E 
(Pucinskas, 1970).
The Sokoloski, Bildsten \& Ho (2001) total observation time was $\simeq$ 12 hr, and  
variability was detected in two runs, with variance twice that expected from noise 
and instrumental effects, but no periodicity has been identified. 
 
\section {High speed photometric observations}

Time resolved photometry was conducted for twenty one nights during four runs
from August to September 2003, on eight runs in July and August 2004 and on six 
short runs in May 2007.  
Their duration varied from less that 1 hr up to $\sim $9 hr. The total observation 
time of BF Cyg is 112.7 hr for the year 2003, 58.45 hr for 2004 and 5.0 hr for the  
year 2007. Altogether, more than  176 hr of time-series data were obtained. 
Table 1 gives a listing of all the runs.

\begin{figure*}
\includegraphics[width=100mm]{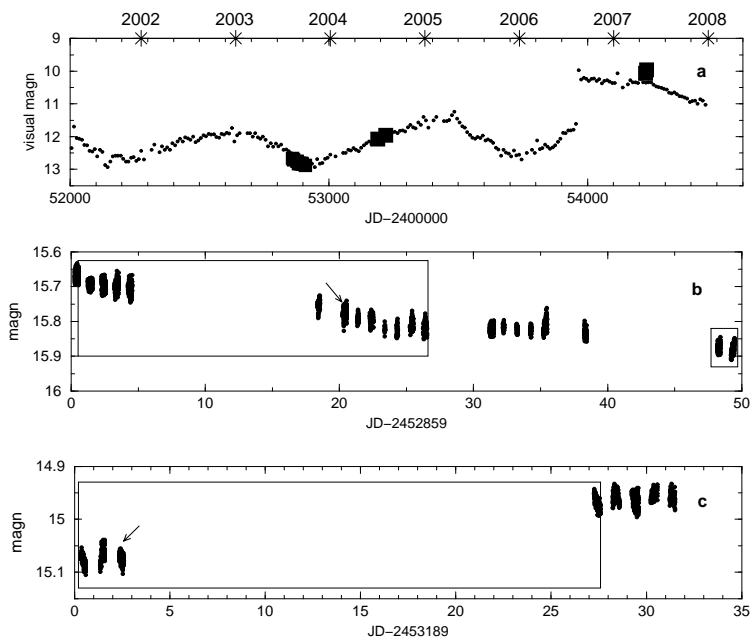}
\caption{ (a) The  LC of BF Cyg: points are 24 days averaged AAVSO magnitudes, filled squares are
our instrumental B magnitudes. (b) The twentyone nights of our observations in the 2003 year. (c) The eight nights of our observations in the 2004 year. The arrows and the rectangles are explained in the text.} 
\end{figure*}

The photometry was carried out with the 1 meter telescope at the Wise Observatory, 
using the Tektronix CCD camera. A Johnson filter B was used for all the observations.
The  exposure time was almost always 120 sec. 
The CCD frames were processed in a standard way for bias removal and 
flat field correction. Aperture photometry was performed using the IRAF DAOPHOT 
procedure. Differential magnitudes have been calculated relative to a set 
of reference stars, using the Wise Observatory reduction program DAOSTAT 
(Netzer et al, 1996). 
The star GSC 02137-00847, discovered to be a pulsating $\delta$ Scuti  
by Sokoloski et al. (2002), was excluded from the set of comparison stars. In fact 
it was outside the field of our frames in most of the observations.

No large fluctuation or flare activity occurred in the system during 
our 2003 and 2004 photometric coverage, while the 2007 data were obtained 
during an outburst epoch. 
Fig. 1 (a) presents the visual LC of BF Cyg obtained by averaging the AAVSO data
over a time interval of 24 d. The instrumental magnitudes of BF Cyg in the B filter, as 
measured at the Wise Observatory in the years 2003, 2004 and 2007, are superposed
on the  AAVSO  light curve, applying an arbitrary shift.  One can see that the 2003 
observations caught the star as its brightness was declining towards minimum light 
in its 757 d binary cycle. 
In 2004 the star was on the ascending branch of its binary photometric cycle and 
it was brighter than in the year before by about 1 mag. By the year 2007 it was  
nearly three magnitudes brighter than in 2003, as the system was undergoing an outburst
activity.

\section{Time Series Analysis}

Fig. 1(b) is zoomed in on the time axis of the 2003 data. The 21 nights of our observations 
in that year, as well as the relative mean magnitude of each night, are well resolved 
in this frame. The arrow points at  one night, the LC of which is shown in Fig. 2(a).
Here, the scale of the time axis allows the display of all individual
measurements at that night.
The rectangles in Fig. 1(b) will be explained in Section 4.1. 
Fig. 1(c) depicts the 2004 LC, with the arrow pointing at the night, the high resolution 
LC of which is shown in Fig. 2(b). An eye inspection of Fig. 2(a) and (b) is enough to 
realize that the star light at these two nights was varying on time scale of one hour 
with a relative amplitude of one percent of its total brightness. 

\subsection{Periodicities}

In order to  further investigate the nature of these variations we computed the power 
spectrum (PS) of the LC of each year (Scargle 1982), within the frequency range 5-40 d$^{-1}$. 
Our procedure consists of pre-whitening the LC by subtracting from each night magnitude 
values, a polynomial of second degree that was fitted to the observed data by least squares. 
This operation is performed in order to remove from the data variations on time scale of one 
night (a few hours) that may be present in the data at the level of one percent magnitude, due 
to temperature variation during the night, stability of the voltage in the CCD camera, 
and to residual color effects in the atmospheric extinction correction applied to 
the measured magnitudes. Heliocentric Julian days were calculated.
Fig. 3 (a) displays the PS of the 2003 LC in the above mentioned frequency interval. 
The highest peak, around frequency f=26.6 d$^{-1}$ is statistically significant at a  
better than 99.9 percent confidence level. The confidence level is the false 
alarm probability calculated according to Scargle (1982) and Horne \& Baliunas (1986).  
This quantity represents the probability that the data contain a  signal with respect 
to a data set of pure white noise. Fig. 3 and Fig. 4 show that in the frequency range 
of our interest, this  is indeed the case, to a good approximation. We also found that 
the noise in the LCs of the reference stars seems to be of a similar character. 
 
\begin{figure}
\includegraphics[width=80mm]{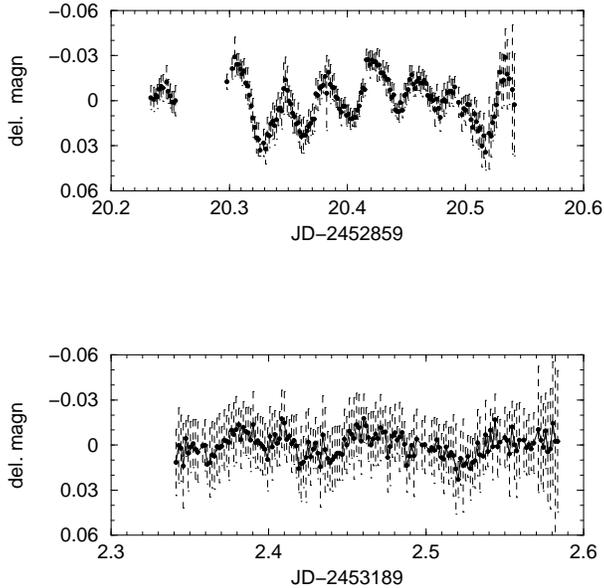}
\caption{ Differential lightcurves for the nights marked by an arrow in Fig. 1 (b) and (c).Vertical lines represent the estimated error range in the displayed magnitudes.}
\end{figure}

\begin{table}
\caption{Observations log  } 
\begin{tabular}{@{}rrlr@{}} \\
\hline
Number&Date &  JD & Obs length (hr)  \\  
     
\hline
1&  Aug   7 2003 &2452859   & 7.33        \\ 
2&  Aug   8 2003 &2452860   & 8.65       \\ 
3&  Aug   9 2003 &2452861   & 7.87        \\ 
4&  Aug  10 2003 &2452862   & 7.59        \\ 
5&  Aug  11 2003 &2452863   & 7.55        \\ 
6&  Aug  25 2003 &2452877   & 4.33       \\ 
7&  Aug  27 2003 &2452879   & 7.45       \\ 
8&  Aug  28 2003 &2452880   & 2.61       \\ 
9&  Aug  29 2003 &2452881   & 6.20        \\ 
10&  Aug 30 2003 &2452882   &  .84        \\ 
11&  Aug 31 2003 &2452883   & 2.56        \\ 
12&  Sept  1 2003 &2452884   & 6.00     \\
13&  Sept  2 2003 &2452885   & 7.18     \\
14&  Sept  7 2003 &2452890   & 7.29     \\
15&  Sept  8 2003 &2452891   & 2.17     \\
16&  Sept  9 2003 &2452892   & 1.98     \\
17&  Sept 10 2003 &2452893   & 2.09    \\
18&  Sept 11 2003 &2452894   & 6.88     \\
19&  Sept 14 2003 &2452897   & 4.26     \\
20&  Sept 24 2003 &2452907   & 5.48    \\
21&  Sept 25 2003 &2452908   & 6.39     \\
\hline
22&  Jul  2  2004 &2453189   & 5.83    \\
23&  Jul  3  2004 &2453190   & 6.02     \\
24&  Jul  4  2004 &2453191   & 6.85     \\
25&  Jul  29 2004 &2453216   & 8.17    \\
26&  Jul  30 2004 &2453217   & 8.13     \\
27&  Jul  31 2004 &2453218   & 8.20     \\
28&  Aug   1 2004 &2453219   & 8.41    \\
29&  Aug   2 2004 &2453220   & 6.84     \\
\hline
30&  May 03  2007 &2454224   & .96     \\
31&  May 04  2007 &2454225   & .95     \\
32&  May 05  2007 &2454226   & 1.38     \\
33&  May 06  2007 &2454227   & 1.44     \\
34&  May 07  2007 &2454228   & .27     \\
\end{tabular}
\end{table}

Inspecting the details of the fit of a Sine wave with the peak periodicity to the observed 
data, we noticed that there are subsections of the LC in which the cycle does not seem to 
represent well the variations in the data points. We have therefore developed a MATLAB 
script for finding out periodicities in subsections of a given LC. The program scans the 
time axis of the LC with a window, the width of which varies from a few cycles of the 
period of the highest peak in the PS of the overall LC, up to a window size that covers 
the entire LC. The window is moved over the time axis in steps of 1 cycle, and the 
program computes the PS of the section of the LC that is contained in each position 
of the window. The program finds the window position at which the highest peak in 
the PS is higher than the peaks in all other positions. We thus have for each window 
size, the frequency and the power of the highest peak, and the section in the LC where 
these values are obtained.

In this way we have discovered that the ~26 d$^{-1}$ oscillations appear in the data 
between two well defined times. The corresponding section of the LC is marked by the 
large rectangle in Fig.1(b).

\begin{figure}
\includegraphics[width=90mm]{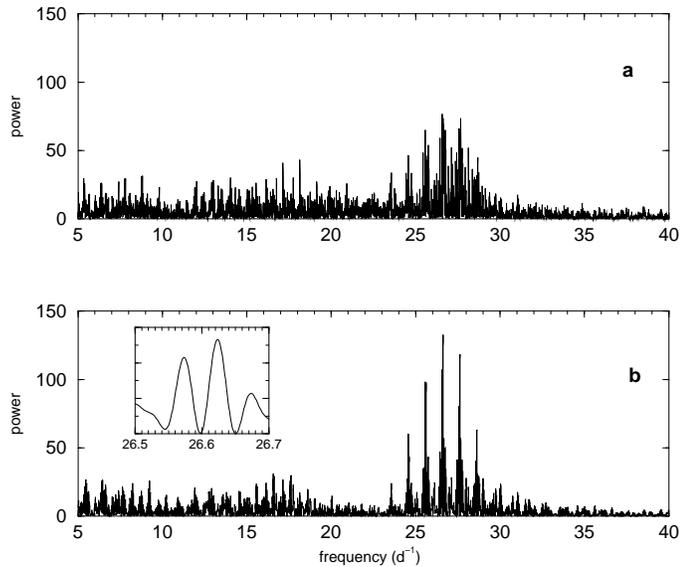}
\caption{(a) Power spectra of the 2003 twenty one  nights of BF Cyg. (b) Power spectra of the 2003 nights inside the rectangles marked in Fig.1(b).}
\end{figure}

\begin{figure}
\includegraphics[width=90mm]{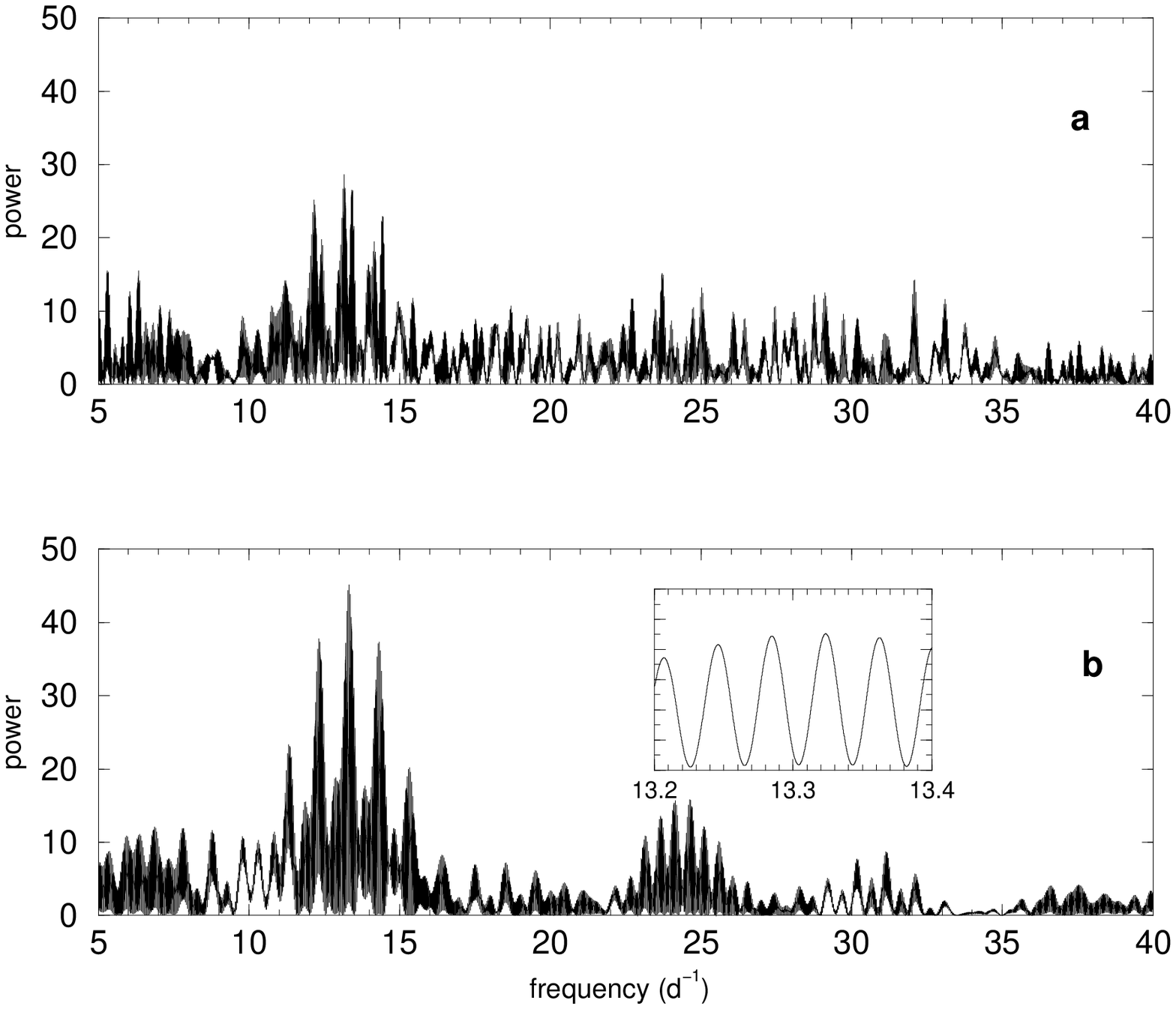}
\caption{(a) Power spectra of the 2004  nights of BF Cyg. (b) Power spectra of the 2004 nights inside the rectangle marked in Fig.1(c).}
\end{figure}

If we extend the subsection under consideration either to the right or to the left of the 
rectangle, the power in the highest peak is reduced. When we consider the nights 
that are out of the rectangle, no significant peak is present in the PS at all.
However, aperiodic flickering is detected in these nights. 
The 26.6  d$^{-1}$ frequency appears again in the LC, in the subsection marked by 
the small rectangle on the right hand side of Fig. 1(b).

Fig. 3(b) depicts the PS of the LC consisting of the two rectangles together. 
The relative strengthening of the periodic component 
in this LC is hardly in need of further elaboration. The satellite peaks on the 
two sides of the main peak are of course due to one day aliases of the cycle true 
frequency. 

We found that when considering first the PS of the data points within the first 
rectangle and then adding up the points in the second rectangle, 
the power of the high peak in the PS grows without changing its position. 
This fact indicates that the phase of the periodicity in the two rectangles  
is the same. In the next subsection we discuss the coherence of the 
periodicity in more quantitative terms.

The insert in Fig. 3(b) is zoomed in on the highest peak of the main graph.
It  shows that the main peak is in fact a doublet. The frequencies of the two 
components are 26.620  d$^{-1}$ and 26.576 d$^{-1}$. The uncertainty in  
these values, as judged by the half width at half maximum of the  
corresponding peaks is $\pm$0.006. We also made a second estimation of 
the uncertainty in the frequency value on the basis of a sample of  
10000 bootstrap pseudo-LCs (Efron and Tibshirani 1993), which gave
us a similar value for a 90 percent uncertainty interval.

We applied a second period search routine, computing by least squares  
the residuals of the observed data points from an harmonic wave  
consisting of the first two harmonics of a given frequency.
Using the $\chi^{2}$ value as a measure of the goodness of fit, and 
scanning the above  mentioned frequency range, we found deep minima in the 
plot of $\chi^{2} $ vs. frequency at the frequencies f'=13.310 d$^{-1}$  
and f=13.288 d$^{-1}$, respectively. Their second harmonics are exactly the 
frequencies of the two highest peaks in the PS of this LC. 
The two frequencies, as well as the frequencies of the corresponding second 
harmonics, are aliases of each other, due to the $\sim$20 days gap between 
the mean times of the two major groups of successive nights of our observations, 
seen within the large rectangle of Fig. 1(b). 
The inter-dependence between these two frequencies is manifested by the fact 
that if one of them is removed from the data, in the PS of the residuals, 
the peak of the other one is being removed as well. For reasons explained below, 
we believe that the major periodicity in the LC of 2003 is the one corresponding 
to the lower of the two peaks seen in the insert of  Fig. 3(b), namely the one of 
the frequency 2f=26.576 d$^{-1}$.
Fig. 4(a) is the PS of the 2004 LC, within the same frequency interval.
The highest peak is concentrated around the frequency 13.31 d$^{-1}$.
Applying again our search script on these data we find that the highest peak in the 
PS is obtained in the subsection of the LC delimited by the rectangle in Fig. 1(c). 
No significant peak is found in the PS of the LC measured in the nights outside 
that rectangle.

 Fig. 4(b) presents the PS of the delimited section of the 2004 LC. 
The highest peak is much more pronounced here and it is clearly very 
significant. Its formal statistical significance of  better than 99 percent 
confidence level is computed as described above for the 2003 LC.
The insert in Fig. 4(b) is zoomed in on the highest peak of the main  graph. It shows 
that this major peak is in fact a multiplet. The two highest components of similar 
peak power are at the frequencies 13.323 d$^{-1}$ and 13.285 d$^{-1}$ with an 
uncertainty of $\pm 0.018$. Here again, the two frequencies are aliases of each other, due to 
the $\sim$26 day interval between the three first nights and the fourth night seen 
in the rectangle in Fig. 2(b).
With our second periodogram routine that finds the frequency of a Sine wave  
whose two first harmonics are fitted simultaneously to the data, we 
obtain the deepest minimum of the $\chi^{2}$ value at the frequency 13.324 d$^{-1}$ 
and a second minimum of nearly equal depth at the frequency 13.286 d$^{-1}$ .
The frequency 13.286 d$^{-1}$ is nearly equal to the f=13.288 d$^{-1}$ frequency, 
that  is found in the independent LC of the year 2003.

We also computed the PS of the 2007 LC of BF Cyg. No significant peak  
is present in that power spectrum.

The analysis presented in this Section implies that a  periodicity with 
the frequency f=13.288 d$^{-1}$ and of twice this frequency was intermittently 
present in the LC of BF Cyg in 2003 as well as in 2004.

\subsection{Phases}

In this Section we turn to the crucial question of how coherent the 13.3 frequency 
was during our two observing seasons.
 We denote by S1 the set of the  nights of our observation in 2003 shown 
within  the large rectangle in Fig. 2(b). The set of the two nights in the other 
rectangle is denoted by S2. The PS of the S1 LC is similar to the PS of the LC 
of the entire 2003 observing season. In particular, the oscillating component 
has the same frequency 2f as that of the entire 2003 LC.

Fig. 5(a), (b) and (c) present the S1, S2 and the 2004 LCs, respectively, folded 
onto the P=1/f periodicity. All three plots refer to the same zero time at 
HJD=2452859.00. The cycle is shown twice. Fig. 5(d) presents the binning of each of 
the folded LC shown in the upper frames into 22 bins. The coincidence of the phases 
of maximum light of the three curves is quite evident in the figure.

The statistical significance of the similarity of the phases of the three LCs may 
be estimated in quantitative terms, by evaluating the probability of a null 
hypothesis that they are independent of one another. 
By least square procedure we compute  the phases of the two first harmonics of 
the frequency f=13.288 d$^{-1}$ fitted simultaneously to the S1 data points.
In a presentation of each harmonic component as  $A \times sin[2\pi(t/P+\phi)] $   
we find the phase value $\phi =0.1291$ for the major, second harmonic of this 
frequency, taking HJD=2452859.00 as time zero. 
In a similar way, we independently compute the phases of these two harmonics, when fitted 
to the S2 data. The phase of the second harmonic in S2 is 0.1454. When considering the
entire 2003 sample, the  second harmonic phase is 0.1322.

We then compute the phases of the two harmonics fitted to the 2004 data. 
Here the phase of the first component is 0.8327.

Consider first the phases of the S1 and the S2 LCs. The difference between them is 
0.0163. For two independent phase values, the probability that they fall at random 
within an interval in phase space of width 0.0163 is p1=0.0323.

Consider now the phase relation between the first and the second  harmonics of a 
Sine wave. Let $\phi_{1}$ be the phase of the f component and  $\phi_{2}$
the phase of the 2f component. The 2f component has two minima (phases of maximum 
light in magnitude units) within one cycle of the f component. The first minimum of 
the 2f cycles coincides with the minimum of the f cycle when the two phases are 
related as follows:  $\phi_{1}-\phi_{2}$/2=3/8. The second minimum of the 2f wave 
coincides with the minimum of the f cycle when $\phi_{1}- \phi_{2}$/2=7/8=0.875 
The  corresponding measured difference between the phases of the 2004 LC 
and that of the 2003 LC  is 0.7666. The difference from the coincidence value 
0.875 is -0.1084.
The probability that the phases of two independent frequencies f and 2f will 
differ from one another, as a random event, by a number that is within a distance 
of  0.1084 from one of the two values of strict coincidence of maximum light is 
$p2=4\times 0.1084=0.4337$.

The probability that the phases of the S1, S2 and the 2004 LCs are  
independent of one another is therefore the product $p1\times p2 =.0140$. 
The null hypothesis that the three phases are independent of one another can 
therefore be  rejected at a 98.6 percent confidence level.

We note that, since the observations of the two years are separated by $\sim 4400 $ 
cycles, the fit of the phases between the two years, is very sensitive to the exact 
value of the frequency. Il we take  f= 13.2877 instead of 13.288, the phases 
of the two years are within 0.011 of the exact fit value. With this value of the frequency, the probability of a random coincidence of the three phases is reduced to a value less than
0.48 percent.

\begin{figure}
\includegraphics[width=80mm]{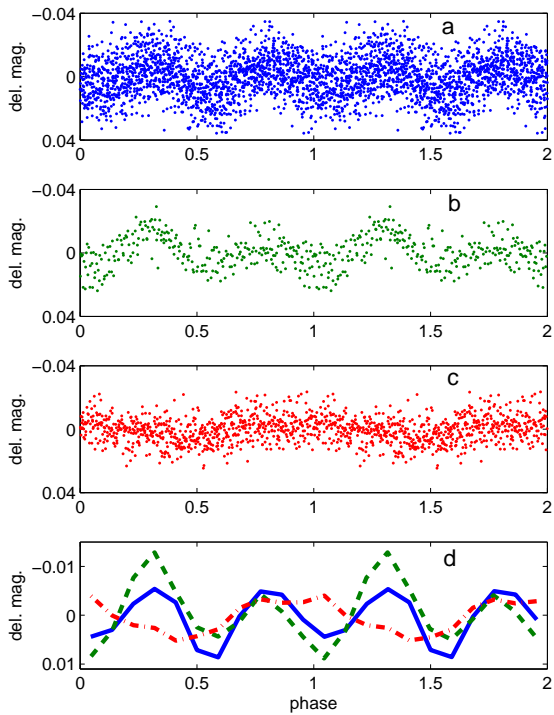}
\caption{(a) The section  S1 of the 2003 light curve folded onto the period corresponding to f=13.288 d$^{-1}$. (b) The section S2  of the 2003 light curve folded onto the same period with the same zero time. (c) The 2004 light curve inside the rectangle folded onto the same period with the same reference time. (d) The three curves display the folded LCs 
in (a), (b) and (c), each one binned into 22 bins. Continous line refers to (a), dashed line to(b), and dot-dashed line to (c).}
\end{figure}

\subsection{Amplitudes}

The amplitude of the cyclic oscillations, in magnitude units, was 0.007 in 2003. 
In 2004 it was 0.004.

\section{Discussion} 

Our time series analysis of the B LC of BF Cyg in 2003, 2004 and 2007 revealed cyclic 
variations in the first two years and none in the third one. Within the rather narrow 
range of uncertainty in the frequency of the oscillations in 2004, its value is one 
half of the frequency measured in 2003. In both years, the cyclic oscillations were 
highly coherent, retaining a constant phasing within each year, and very likely also 
between the two consecutive years. The stability of the phase is particularly remarkable 
in view of the fact that the cyclic variations themselves were present in the LC only 
in some fractions of the entire LC, while being absent from some other subsections of 
the measured LC. In the year 2007, when the star was three magnitudes brighter than in 2003, 
no variability above the observational noise and  an particular no cyclic variations 
seem to be present at all in the LC.

The coincidence of the frequency f(2003) with twice the value f(2004) leads us to conclude 
that the oscillations in the two years are of the same fundamental period, with a two 
hump cycle in 2003 and one hump cycle in 2004. The period is  P=1.806 hr. 
The high degree of coherence of the cyclic oscillations, well indicated by the conservation 
of the phase over thousands of cycles between the two years 2003 and 2004, requires a 
driving clock of very high precision. The spin of the hot component of the binary 
system of BF Cyg, the WD star, suggests itself as this time keeping clock.

We hypothesize that the cyclically oscillating component in the light of BF Cyg originates 
at one or two antipodal regions on the surface of the rotating WD star. These could be 
areas around the two opposite poles of an oblique dipole magnetic field of the star. 
The two hot spots are created by accretion of matter from the wind of the giant star 
of the system. The matter is funneled by the WD magnetic field to the areas near the poles. 
The hot zones are on the surface of the star or at the top of accretion columns, where 
the falling material hits the ground. 

The accretion flow in 2003 was two headed, with 
matter directed to the two poles of the magnetic field, while in 2004 matter was 
impinging the ground mainly around  only one of the poles. It seems, however, that 
the accretion process itself was also taking place between the episodes of coherent 
oscillations as indicated by the a-periodic variability of the LC  during the 
time intervals outside the rectangles in Fig. 1(b) and (c). 
  
In Section 4 we reported that the amplitude of the cyclic oscillations in 2003 
was 0.007 mag, while in 2004 it was 0.004 mag. In 2004, the star brightness 
in the B filter was about twice the value in the previous year, see Fig. 1(a). 
These numbers are consistent with the notion that the absolute brightness of the 
hot spots  does not change much in time. 
The decline in the amplitude of the oscillations is due entirely to the brightening of 
the background DC luminosity of the system. This may also well explain the reduction of 
the amplitude of the variability of the LC of the year 2007 to the level of the noise 
in the measured magnitude values. In this year the system was some three magnitudes 
brighter than in 2003. The background luminosity reduces the relative brightness of the 
varying component to below our detection ability.

A detailed theoretical model of the dynamics of the accretion, as well as further 
observations, are obviously required in order to substantiate the qualitative scenario 
that we are suggesting here.

\section {Conclusions}
This work is a report of the discovery and measurement of the spin period of the WD in 
the symbiotic stellar system BF Cyg.  The period is found to be {\bf 1.806 hr.}
BF Cyg is the second SS system for which a stable periodic oscillation was 
found thus far. The other  symbiotic with known WD spin periods is
Z And, with a WD spin period 28 min (Sokoloski \& Bildsten 1999).
For Z And, the serendipitous discovery occurred in a particular state of the 
system, during the decrease from a small outburst, while the oscillations disappeared 
near optical maximun (Sokoloski et al. 2006).

We believe that the periodic signal already detected for Z And and BF Cyg is not 
unique to these particular systems. The long temporal baseline of our observations 
and the fact that the data were obtained in many consecutive nights enabled us to 
discover the phenomenon.

It also seems that the total brightness and/or the binary phase of
the system at the time of the observations are critical for the detection of 
a periodic signal in the light curve.

With just two SS systems with known WD spin period, one can hardly arrive at a meaningful 
conclusions about the significance of the value of this parameter for our understanding 
the evolutionary stage that symbiotics are  in the history of binary stars. 
It may be of significance that in the two known cases, the spin period of the WD is of 
the order of an hour. 
When a larger statistics is assembled, it would be of interest, for example, to compare 
it with the spin period of the WD in another class of highly interacting  binary systems,
namely the family of dwarf and classical novae. 

\section*{Acknowledgments}

We acknowledge the anonymous referee for  his/her helpful suggestions.
We are grateful to Yiftah Lipkin for his help in collecting the observations
and to Margie Goss for her careful reading of the manuscript.
We acknowledge with thanks the variable star observations from the AAVSO
International Database, contributed by observers worldwide and used in this
research. 
This research is supported by ISF - Israel Science Foundation of the
Israeli Academy of Sciences.

\label{lastpage}

\begin{thebibliography}{}
\bibitem[]{} Dobrzycka D., Kenyon S.J, Milone A.A.E., 1966, AJ, 11, 414
\bibitem[]{} Efron B., Tibshirani R.J., 1993, An Introduction to the Bootstrap, Chapman \& Hall, New York, London
\bibitem[]{} Fernandez-Castro T., Gonz\'{a}les-Riestra R., Cassatella A., Fuensalida J.J., 1990, A\&A, 227, 422
\bibitem[]{} Formiggini L., Leibowitz E.M., 1994, A\&A, 292, 534
\bibitem[]{} Formiggini L., Leibowitz E.M., 2006, MNRAS, 372, 1325 
\bibitem[]{} Gonz\'{a}les-Riestra R., Cassatella A., Fernandez-Castro T., 1990, A\&A, 237, 385
\bibitem[]{} Gromadzki M., Mikolajewski M., Tomov T., Bellas-Velidis I., Dapergolas A., Galan C., Acta Astron, 56,97
\bibitem[]{} Horne J.H, Baliunas S.L., 1986, ApJ, 302, 757
\bibitem[]{} Kenyon S.J., The symbiotic stars, Cambridge Univ. Press., 1986
\bibitem[]{} Leibowitz E.M., Formiggini L., 2006, MNRAS, 366, 675 (paper I)
\bibitem[]{} Leibowitz E.M., Formiggini L., 2008, MNRAS, 385, 445 
\bibitem[]{} Mikolajewska J., Kenyon, S.J., Mikolajewski M., 1989, AJ, 98, 1427
\bibitem[]{} Munari U., Siviero A., Moretti S., Graziani M., Tomaselli S., Baldinelli L., Maitan A., 2006, CBET, 596
\bibitem[]{} Netzer et al., 1996, MNRAS, 279, 429
\bibitem[]{} Pucinskas A., 1970 Bull. Vilnius Univ. Astron. Obs. No.27, 24
\bibitem[]{} Scargle J.D., 1982, ApJ., 263, 835
\bibitem[]{} Skopal A., Vanko M., Pribulla T., Chochol D., Semkov E., Wolf M., Jones A. 2007, AN 328,909
\bibitem[]{} Sokoloski J. L., Bildsten, L., 1999, ApJ,517, 919
\bibitem[]{} Sokoloski J. L., Bildsten, L., Ho, W.C.G.  2001, ApJ, 326, 553
\bibitem[]{} Sokoloski J.L., Bildsten L., Chornock R., Filippenko A.V. 2002, PASP, 114, 636
\bibitem[]{} Sokoloski J.L., Kenyon S.J 2003, ApJ, 584, 1021
\bibitem[]{} Sokoloski J. L. et al., 2006, ApJ, 636, 1002 
\end{thebibliography}
\end{document}